\documentstyle[12pt]{article}
\advance\oddsidemargin	by -1in
\advance\evensidemargin by -1in
\advance\topmargin	by -0.0in
\makeindex
\newcommand\beq{\begin{equation}}
\newcommand\eeq{\end{equation}}
\newcommand\beqn{\begin{eqnarray}}
\newcommand\eeqn{\end{eqnarray}}
\newcommand{\doublespace} {
    \renewcommand{\baselinestretch} {1.6} \large\normalsize}

\begin{document}

\vspace*{0.5cm}
\hspace*{9cm}{\Large\bf MPIH--V11--1995}
\vspace*{1.5cm}

\begin{center}

\centerline{{\huge \bf
The Relative $J/\Psi$ to $\Psi'$ Suppression}}
\vspace{0.4cm}
\centerline{{\huge \bf in Proton-Nucleus and}}
\vspace{0.5cm}
\centerline{{\huge \bf Nucleus-Nucleus Collisions}}

\vspace{1.5cm}

\bigskip

 {\Large J\"org~H\"ufner}\\
\medskip
{\sl Instut f\"ur Theoretische Physik der Universit\"at ,\\
 Philosophenweg
19, 69120 Heidelberg, Germany\\
 E-mail: huefner@zooey.mpi-hd.mpg.de}
\bigskip

 {\Large Boris~Kopeliovich}\\
\medskip
{\sl Max-Planck Institut f\"ur Kernphysik\\ Postfach 103980,
 69029
Heidelberg, Germany}\\

 and\\
 {\sl Joint Institute for Nuclear Research,
Laboratory of
 Nuclear Problems,\\
 Dubna, 141980 Moscow Region, Russia\\
E-mail: bzk@dxnhd1.mpi-hd.mpg.de}\\

\end{center}
\bigskip

\begin{abstract}
We calculate the nuclear suppression for $J/\Psi$ and $\Psi'$
 production
within a coupled channel approach in the subspace of
 the $J/\Psi$ and $\Psi'$
states.  We are able to explain, why (i)
 the $J/\Psi$ and $\Psi'$ show the
same suppression from $200\
 GeV$ to $800\ GeV$ in proton-nucleus collisions
and why (ii) the
 $\Psi'$ is absorbed more strongly than the $J/\Psi$ in
nucleus-nucleus collisions at $200\ GeV$. The numerical result
 which includes
only interactions with nucleons acconts for half
 of the observed suppression
in sulphur-uranium collisions.

\end{abstract}

\newpage
\doublespace

 The E772 collaboration \cite{e772} was the first who claimed
 that $\Psi$
(we use hereafter this abbreviation for $J/\Psi$) and
 $\Psi'$ produced in
proton-nucleus ($p-A$) collisions at $800\
 GeV$ and $x_F\geq 0.1$ experience
the {\sl same} nuclear
 suppression.  This result was confirmed by the NA38
collaboration
\cite{na38'1,na38'2,na38'3} at an energy of $200\ GeV$ and for
$x_F\approx 0.1$, though with larger error bars.  The two
 observations
contradict the simple-minded expectation, namely
 that the $\Psi'$ should be
more strongly suppressed, because
 absorption cross section scales with the
mean square radius
 $\langle r^2\rangle$ of the meson \cite{gs,hp} and the
$\Psi'$ is
 much larger than the $\Psi$.  To our opinion, no satisfactory
explanation of the data has yet been proposed.

 Recently the situation
became even more mysterious by the
 observation \cite{na38'1,na38'3} that in
Sulfur-Uranium (S-U)
 collisions at $200\ GeV/A$ the $\Psi'$ is significantly
more
 strongly absorbed than the $\Psi$.  Does this result signal dense
hadronic gas or quark-gluon plasma formation?

 In this letter we treat the
propagation of a $c\bar c$ pair
 through nuclear matter as a {\sl coupled}
system of the $\Psi$
 and $\Psi'$ states. In addition to elastic collisions
$\Psi N\ss\Psi N$ and $\Psi' N\ss\Psi' N$ with amplitudes
 $f(\Psi,\Psi)$ and
$f(\Psi',\Psi')$ respectively, we consider the
 conversion amplitudes
$f(\Psi,\Psi')$ and $f(\Psi',\Psi)$ for the
 processes $\Psi
N\rightleftharpoons\Psi'N$ during propagation
 through nuclear matter. The
inelastic amplitudes turn out to be
 nearly as big as the elastic ones.

We define as nuclear suppression factor of a $\Psi$ meson in $pA$
 collisions
the ratio

\beq
S^{pA}_{\Psi}(x_F,E)={1\over
 A}\frac{\sigma(pA\ss\Psi X;x_F,E)/dx_F}
{\sigma(pN\ss\Psi X;x_F,E)/dx_F}\ ,
\label{1}
\eeq
where $E$ is the lab. energy. We also introduce the relative
 $\Psi'$ to
$\Psi$ nuclear suppression function by

\beq
S^{pA}_{\Psi'/\Psi}=
 S^{pA}_{\Psi'}/
 S^{pA}_{\Psi}
\label{2}
\eeq

 Then the results of the E772 and NA38 experiments
 can be summarized by
$S^{pA}_{\Psi'/\Psi}\approx 1$ for the
 values of $x_F$ and $E$ considered.

 In order to exhibit the physics of our coupled channel approach,
 we first
calculate eq. (\ref{2}) perturbatively, i.e.
 restricting ourselves to only
{\it one} $\Psi N$ ($\Psi' N$)
 interaction after the creation of the $c\bar
c$ pair. This
 interaction can be an elastic one as well as a conversion
event.
 Then

\beq
S^{pA}_{\Psi'/\Psi}=\frac{
 1-{1\over 2}\sigma_{tot}^{\Psi N}
(r+\epsilon/R)\langle T\rangle_A}
 {1-{1\over 2}\sigma_{tot}^{\Psi N}
(1+\epsilon R)\langle T\rangle_A}\ ,
\label{3}
\eeq
where $r=f(\Psi'\Psi')/f(\Psi\Psi)$, $\epsilon=
 f(\Psi\Psi')f(\Psi\Psi)$ and
$\sigma^{\Psi N}_{tot}=2Im
 f(\Psi,\Psi)$. Furthermore, $R$ denotes the
relative amplitude of
 $\Psi'$ to $\Psi$ production in the initial $pN$
collision, and
 $\langle T\rangle_A=1/A\int d^2bT^2(b)$, where
$T(b)=\int_{-\infty}^{\infty}dz\rho_A(b,z)$ is the nuclear
 thickness.

 If
one neglects the conversion rate ($\epsilon=0$), eq. (\ref{3})
 reduces to the
conventional result, namely that the $\Psi$
 suppression depends only on
$\sigma^{\Psi N}_{tot}$, while
 $r\sigma^{\Psi N}_{tot}=
\sigma^{\Psi' N}_{tot}$ determines the suppression of the
$\Psi'$.

 The ratios $r$ and $\epsilon$ in eq. (\ref{3}) can be fairly
reliably calculated since at high energies the scattering
 amplitudes
$f(\alpha,\beta)$, where $\alpha$ and $\beta$ stand
 for $\Psi$ and $\Psi'$,
are proportional to
 $\langle\alpha|r^2_T|\beta\rangle$, as long as
 the
meson radius in the
 transverse direction,
 $\langle
 r_T^2\rangle$, is
small.
 With $1S$ and $2S$ harmonic oscillation
 functions for $\Psi$ and
$\Psi'$, respectively, one has $r=7/3$
 and $\epsilon=-\sqrt{2/3}$. The ratio
$R$ of initial $\Psi$ to
 $\Psi'$ production cannot be calculated in this
way. We turn the
 argument around and calculate from eq. (\ref{3}) that value
of
 $R_{cal}$ which leads to the observed relative suppression
$S^{pA}_{\Psi'/\Psi}=1$ and compare it to $R_{exp}$ deduced from
 $pN$
collisions. A quadratic equation for $R$ leads to
 $R_{cal}=\sqrt{5/3}\pm
\sqrt{2/3}$, the smaller one being
 $R_{cal}=0.47$. If one uses the
experimental intensities of
 $\Psi'$ and $\Psi$ produced in $pp$ collisions
and corrects them
 for the feedings $\Psi'\ss\Psi$ and $\chi\ss\Psi$, one
arrives at
 an amplitude ratio $|R_{exp}|=0.48\pm 0.06$ \cite{e705,na38'3},
which agrees well
 with the calculated one, indicating that the initially
produced
 state $|\Phi_i^{c\bar
c}\rangle=(|\Psi\rangle+R|\Psi'\rangle)/\sqrt{1+R^2}$ is such as
 to lead to
the same final state attenuation for $\Psi$ and
 $\Psi'$. Mathematically
spoken, $|\Phi^{c\bar c}_i\rangle$ is an
 eigenstate of the final state
interaction matrix,

\beq
\hat f = \left(
\begin{array}{cc}
1 &
\epsilon\\
\epsilon & r
\end{array}\right)
f(\Psi,\Psi)\ .
\label{4}
\eeq

 The property of $|\Phi^{c\bar c}_i\rangle$ being eigenstate of
 $\hat f$ is
equivalent to the statement that $|\Phi^{c\bar
 c}_i\rangle$ has an extreme
value for its transverse size:

\beq
\frac{d}{dR}\langle\Phi^{c\bar c}_i|r^2_T|\Phi^{c\bar
c}_i\rangle=0\ .
\label{5}
\eeq

 The experimental value $|R_{exp}|$ selects the physical state as
 that with
{\it minimal} transverse extension.

 Strictly speaking the result
eq. (\ref{3}) is only valid for
 energies $E\ss\infty$. For finite
lab. energies, especially for
 $pA$ and $AA$ collisions at $200\ GeV$ one has
to include the
 effect of the longitudinal momentum transfer $q$ associated
with
 the conversion reaction $\Psi\rightleftharpoons\Psi'$:

\beq
q=\frac{M_{\Psi'}^2-M_{\Psi}^2}{2Ex_1}\ ,
\label{6}
\eeq
where $x_1=(x_F+\sqrt{x_F^2+4M_{\Psi}^2/s})/2$ with $\sqrt{s}$
 the
c.m. energy. One arrives at an expression like eq. (\ref{3})
 where $\epsilon$
is replaced by $\epsilon F_A(q)$ with
\beq
F_A(q)=\frac{2}{A\langle T\rangle}
\int_{-\infty}^{\infty}dz\rho_A(b,z)\int_{z}^{\infty}dz'
\rho_A(b,z')\exp(iqz')
\label{7}
\eeq
being a kind of nuclear formfactor. For the E772 experiment
 $q\leq 0.06\
fm^{-1}$ for $x_F\geq 0.2$ and $F_A=1$ is a very
 good approximation. Thus,
the observed result
 $S_{\Psi'/\Psi}^{pA}=1$ is reproduced.

 For the NA38
experiment at $200\ GeV/A$ especially for the
 nucleus--nucleus collisions,
the effect of the formfactor becomes
 crucial: In the $S-U$ collision the
produced $\Psi$ (or $\Psi'$)
 moves with fractional momentum $x_F$ (in the
c.m. system) with
 respect
 to the target nucleus $U$, but moves with $-x_F$
relative to the
 projectile nucleus $S$ (inverse kinematics).
 Therefore the
$\Psi'/\Psi$ suppression arises from
 two sources,

\beq
S^{SU}_{\Psi'/\Psi}=
 S^{pS}_{\Psi'/\Psi}(-x_F,E)
S^{pU}_{\Psi'/\Psi}(x_F,E)\ .
\label{8}
\eeq
For $200\ GeV$ and $x_F=0.2$ we have $q=0.16\ fm^{-1}$ to be used
 in the
second factor of eq.  (\ref{8}) and $q=0.56\ fm^{-1}$ in
 the first one. This
is the inverse kinematics which leads to a
 much stronger $\Psi'/\Psi$
suppression.

 For the detailed comparison with experiment we have to go
beyond
 the perturbative expression (\ref{3}) and use numerical methods.
 We
describe the propagation of $c\bar c$ pair through nuclear
 matter by a
differential equation \cite{kl,jk}

\beq
i\frac{d}{dz}|\Phi^{c\bar c}(z,\vec b)\rangle=
\widehat U(z,\vec b)|\Phi^{c\bar c}(z,\vec b)\rangle\ ,
\label{9}
\eeq
where the limitation to the $\Psi, \Psi'$ subspace invites the
 use of matrix
notations: $|\Phi^{c\bar c}\rangle={\alpha\choose
 {\beta}}$ and

\beq
\widehat U=\left(\begin{array}{cc}0&0\\0&q\end{array}\right)-
{i\over 2}\sigma^{\Psi N}_{tot}\rho_A(\vec b,z)
\left(\begin{array}{cc}1&\epsilon\\\epsilon&r
\end{array}\right)
\label{10}
\eeq
with initial condition $|\Phi^{c\bar
c}_{in}\rangle={1\choose{R}}/\sqrt{1+R^2}$ at the point $(\vec
 b,z_0)$ of
$c\bar c$ creation. At $z=+\infty$ the wave
 function $|\Phi^{c\bar c}\rangle$
is projected on the $\Psi$ and
 $\Psi'$ states.  The result is squared and
averaged over the
 coordinates $(\vec b,z_0)$ of the production point.  Note
that
 this is only true if the longitudinal momentum transfer at the
production point, $q_0=M_{\Psi}^2/2Ex_1\gg q$, is much larger
 than the
reversed mean internucleon distance in a nucleus.
 Otherwise one should take
into account coherence between
 different production points as well (see
discussion
 in terms of production and formation times in \cite{bm,kz}).
This would be equivalent to an effective increase of the length
 of path of
the $q\bar q$ pair in nuclear matter.  However it
 does not affect the
relative $\Psi'/\Psi$ production rate if
 $|\Phi^{c\bar c}\rangle$ is an
eigenstate of interaction.

 The relative $\Psi'/\Psi$ suppressions are
calculated for $p-A$
 collisions and with the help of eq.  (\ref{8}) also for
$A-A$
 collisions.  The numerical results shown in Figs.  1 and 2 are
calculated with realistic nuclear densities and with the values
 of $\epsilon$
and $r$ as given by the harmonic oscillator model
 and for the absolute value
of the $\Psi N$ total cross section
 $\sigma^{\Psi N}_{tot}\approx C\langle
r_T^2\rangle_{\Psi}= 5.7\
 mb$, where we use the perturbative QCD estimate of
\cite{kz} or
 systematics of \cite{hp} for a value of $C$.  While at $800\
GeV$
 the values for $S^{pA}_{\Psi'/\Psi}$ are measured directly, the
 values
given at $200\ GeV$ for
 $B_{\Psi'}\sigma_{\Psi'}/B_{\Psi}\sigma_{\Psi}$ in
nuclear
 collisions were renormalized by us using the value $1.80\pm 0.10\
\%$ for this quantity from $pp$ and $pd$ collisions
\cite{na38'3}.  Fig.  1 shows the suppression functions
$S_{\Psi'/\Psi}$ for $p-W$ at $800\ GeV$ and for $p-U$ and $p-W$
 at $200\
GeV$. Although we predict for $200\ GeV$ some reduction
 of the $\Psi'/\Psi$
relative suppression it is still in agreement
 with the data of the NA38
experiment within rather large error
 bars.  Fig.  2 shows the predicted and
observed
\cite{na38'1,na38'3} suppression $S^{SU}_{\Psi'/\Psi}$ for the
nucleus-nucleus case.  The solid curve is the product of the
 suppression
curves for the $p-U$ and the $S-p$ collisions (each
 dashed).  Experiment and
calculation agree in that the $\Psi'$
 should be significantly more strongly
suppressed than the $\Psi$.
 However, the measured suppression seems to be
stronger than our
 expectation, indicating that there may be room for other
effects.
 The NA38 group has published \cite{na38'1,na38'3} also four
 points
for the relative $\Psi'$ to $\Psi$ suppression in $S-U$
 collisions as a
function of the transverse energy $E_T$.  The
 suppression is larger for
larger values of $E_T$, corresponding
 to more central collisions.  On the
basis of our model we expect
 such a behaviour, but we lack precise
quantitative information
 for the association of a value of $E_T$ to a
definite geometric
 configuration.

 In Fig. 3 we show the $\Psi'/\Psi$
suppression as a function of
 $x_F$ for $Au-Au$ collisions calculated for RHIC
and LHC
 energies. The result is predicted to be energy independent
 at high
energies and the two curves coincide.

\medskip

 The model of coupled channels presented in this letter
 "naturally"
explains, why at high energies of the charmonia
 $\Psi$ and $\Psi'$ should be
similarly suppressed as observed in
 proton-nucleus collisions at $800\ GeV$,
and why in
 nucleus-nucleus collisions one should see significant
differences
 in $\Psi'$ and $\Psi$ suppressions as was reported at $200\
GeV/A$.

 While the model is free of adjustable parameters, the precision
of the two-channel approximation is questionable.
 In order to evaluate the
corrections to $S_{\Psi'/\Psi}$ from
 inclusion of higher charmonium
excitations we switch from the
 hadronic basis to the quark one in coordinate
representation.
 In this case the nuclear suppression of $\Psi$ production
is
 calculated as \cite{kz}

\beq
S_{\Psi}^{pA}=\frac{\left\langle\left|\langle\Phi_{\Psi}^{c\bar
 c}| \widehat
V(b,z,r_T)|\Phi_{in}^{c\bar
 c}\rangle\right|^2\right\rangle}
{\left|\langle\Phi_{\Psi}|
\Phi_{in}^{c\bar c}\rangle\right|^2}
\label{11}
\eeq
and the same for $\Psi'$.  Here the evolution operator $\widehat
 V$ at high
energy (when the fluctuations in $r_T$ are frozen by
 Lorentz time dilation)
reads, $\widehat V(b,z,r_T)=\exp[-Cr_T^2/2
\int_z^{\infty}dz'\rho_A(b,z')]$.  The averaging $\langle
...\rangle_A$ in eq. (\ref{11})
 denotes the integration over the coordinates
of the production
 point weighted with nuclear density.

 Eq.  (\ref{11})
(valid only for $q=0$) generalizes the
 two-channel approach in that
$|\Phi^{c\bar c}_{in}\rangle$ may
 have other components in addition to the
$\Psi$ and $\Psi'$
 states (see an alternative interpretation of enhancement
of the
 $\Psi'$ production rate on nuclei in \cite{kz,bkmnz}).  We have
evaluated the suppression eq.  (\ref{11}) for $\Psi$ and $\Psi'$
 and the
relative suppression $S^{pA}_{\Psi'/\Psi}$ for various
 trial functions for
$\langle\vec r|\Phi^{c\bar c}_{in}\rangle$
 like $\exp(-\alpha r^2)$,
$r_T^2\exp(-\beta r^2)$,
 $r_T^2K_0(\lambda r_T)$ ($K_0$ is a modified Bessel
function),
 where the constants $\alpha$, $\beta$, $\lambda$ have been
chosen
 to give the same amplitude ratio $R$ for the $\Psi'$ content of
$|\Phi^{c\bar c}_{in}\rangle$ relative to the $\Psi$ one.  The
 resulting
relative $\Psi'/\Psi$ nuclear suppression exceeds the
 prediction of the
two-channel model $S_{\Psi'/\Psi}^{pA}=1$ only
 by $5-10\%$ for heavy nuclei,
a deviation which is still
 compatible with the data. Another correction of
the same order of
 magnitude is expected arising from the $\chi$-component of
the
 initial $c\bar c$ state. A more complete analysis of these
 effects as
well as recalculation of the $\Psi'/\Psi$ nuclear
 suppression using
path-integral methods \cite{kz} will be
 presented in a forthcoming paper.

 Note that the initially produced $c\bar c$ wave packet, which is a
combination of $\Psi$ and $\Psi'$, attenuates in the nucleus less
 than each
of two charmonia. This fact is important for the
 nuclear suppression of
$\Psi$, taken separately. This should be
 checked with available experimental
data.

 In the case of nucleus-nucleus collisions, we have assumed that
charmonia attenuate only due to interaction with the projectile
 nucleons,
having $x_F=\pm 1$.  However, particle production (and
 possibly a dense
hadronic gas or a quark-qluon plasma) should
 cause an additional suppression
of $S^{A-A}_{\Psi'/\Psi}$ down
 from our prediction and may explain the
deviation of our
 calculations from the results of the NA38 experiment
depicted in
 Fig.  2.  Recently, Satz \cite{satz} has proposed to use the
value of $S_{\Psi'/\Psi}$ as a probe for dense matter formation.

\medskip

 {\bf Acknowledgement:} We are grateful to P.~Giubellino and
 H.~Satz, who
stimulated our interest to the problem under
 discussion. We thank
Dr. C.~Gerschel for several discussions on
 the experimental
situation. B.K. thanks E.~Predazzi for useful
 discussion and MPI f\"ur
Kernphysik, Heidelberg, for financial
 support. The work was partially
supported by a grant from the
 BMFT, grant number 06HD742(0).

\bigskip

 {\bf Figure capture}

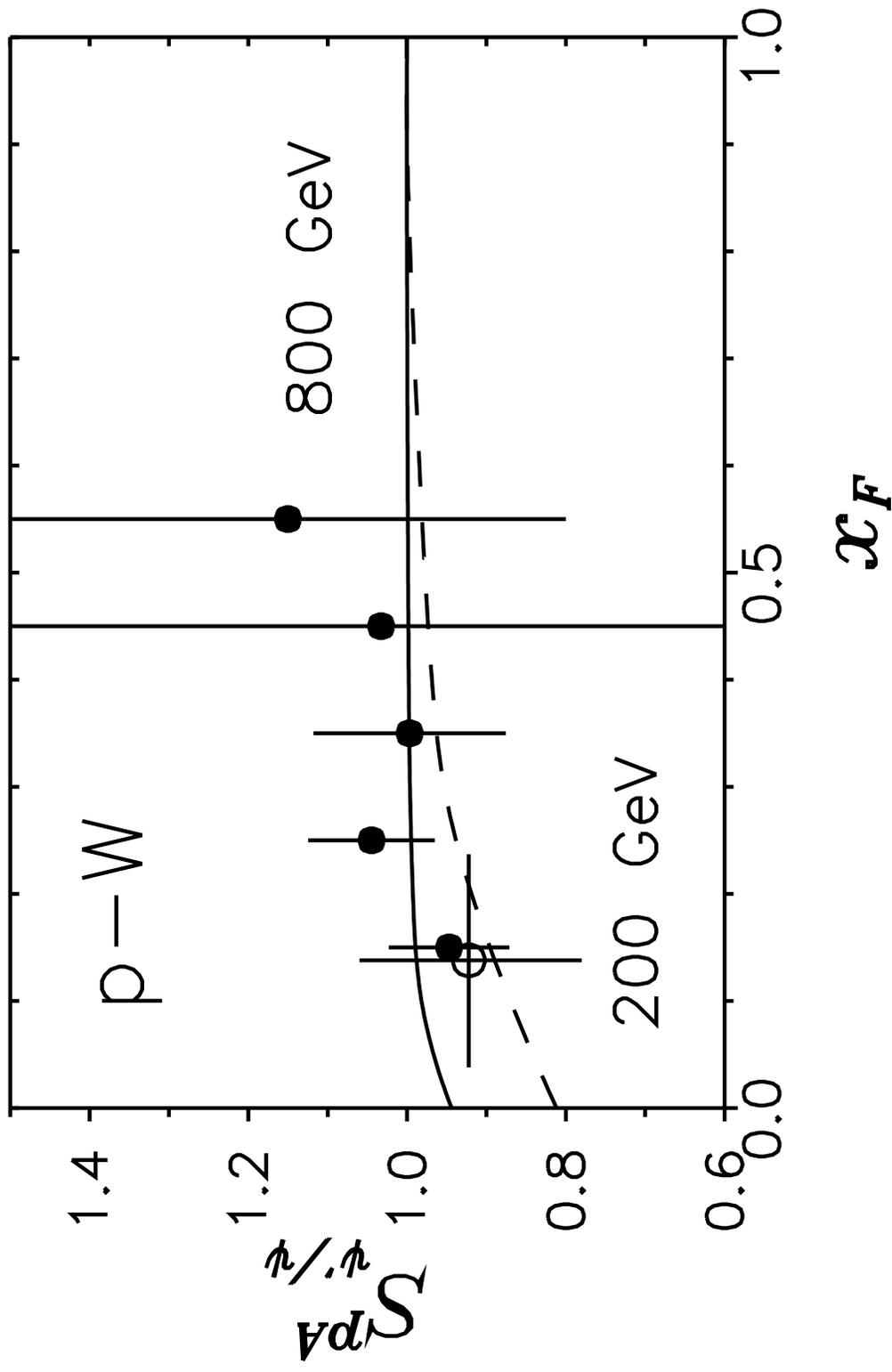
 {\bf Fig. 1} The relative $\Psi'/\Psi$ nuclear
suppression in
 $p-W$ collisions. The full circles and the solid curve are
the
 data of the E772 experiment \cite{e772} at $800\ GeV$ and our
calculation, respectively. The open circle and the dashed curve
 are the
result of the NA38 experiment \cite{na38'3} at $200\ GeV$
 and our prediction,
respectively.

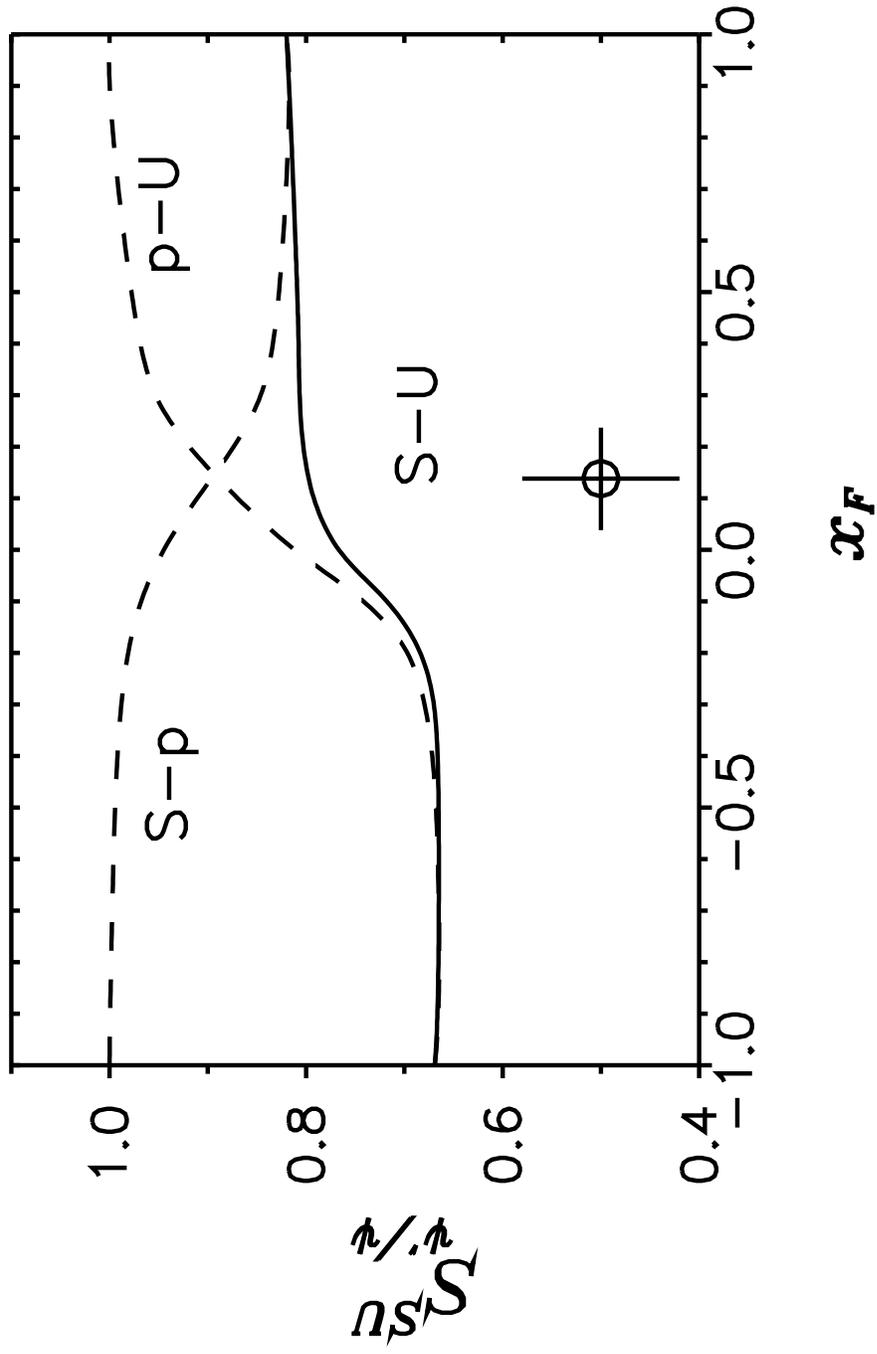
 {\bf Fig. 2}
 The relative $\Psi'/\Psi$ nuclear suppression
in $S-U$ collisions
 at $200/ GeV$. The dashed curves are the relative
$\Psi'/\Psi$
 suppression in $p-U$ and $S-p$ (inverse kinematics) collisions
at
 $200\ GeV$.  The solid curve, which is the two dashed lines
 describes
our prediction for $S-U$ collisions. The data-point is
 the result of NA38
experiment \cite{na38'3}.

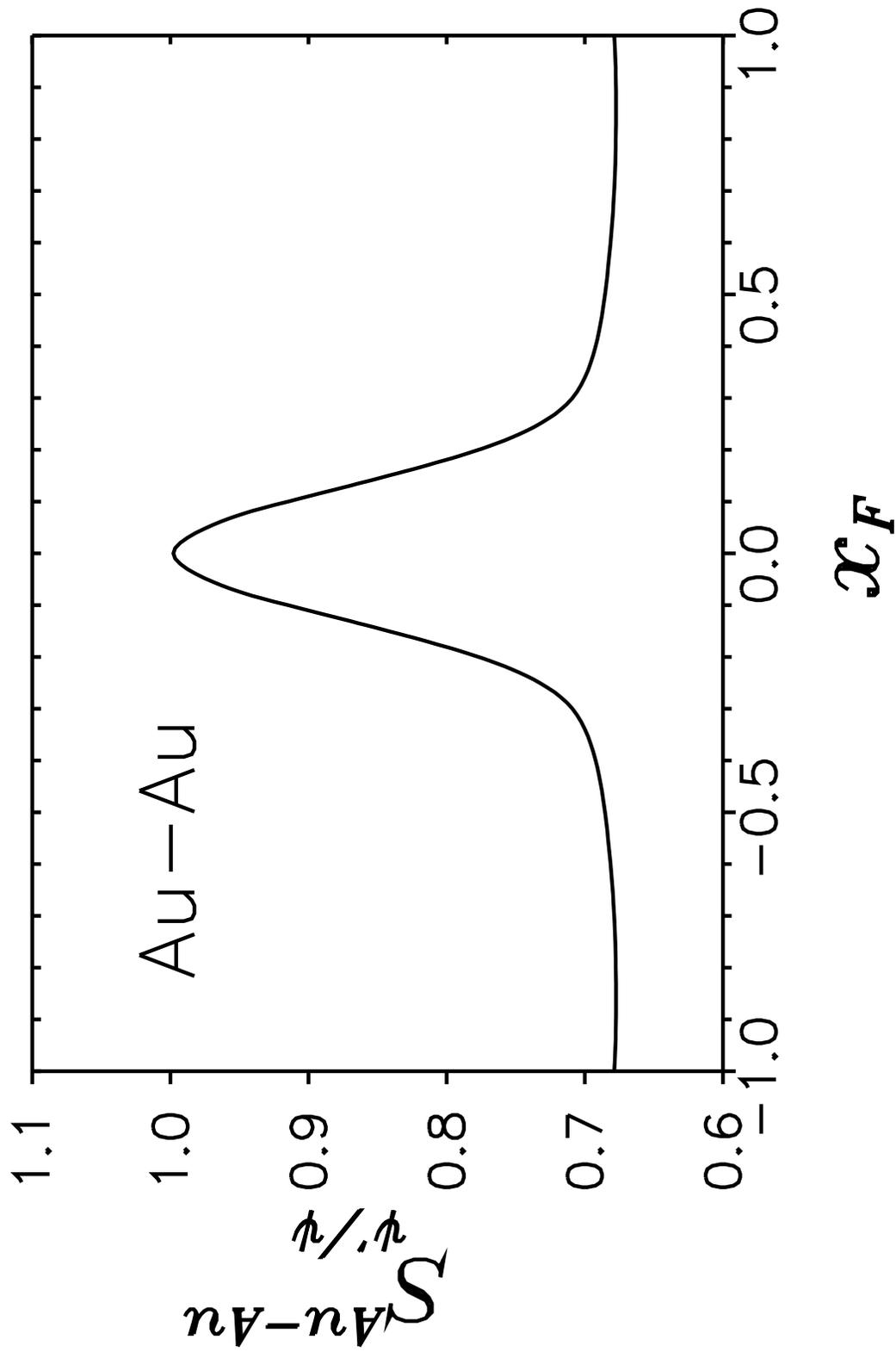
 {\bf Fig. 3} Prediction for the relative
$\Psi'/\Psi$
 suppression in $Au-Au$ collisions for the expected energies
 of
the RHIC and LHC accelerators.


\begin{thebibliography}{MMM}
\bibitem{e772} D.M.~Adle et al., Phys. Rev. Lett. {\bf 66} (1991)
133
\bibitem{na38'1} The NA38 Collaboration, B.~Ronceux et al., Nucl.
Phys. {\bf A566} (1994) 371c
\bibitem{na38'2} The NA38 Collaboration, C.~Lourenco et
al., Nucl. Phys. {\bf A566} (1994) 77c
\bibitem{na38'3} The NA38 Collaboration, C.~Baglin et al., Phys.
Lett. {\bf B345} (1995) 617 and M.C.~Abreu et al., presented at
 QUARK
MATTER'95
\bibitem{gs} J.F.~Gunion and D.~Soper, Phys. Rev.
{\bf D15} (1977) 2617
\bibitem{hp} J.~H\"ufner and B.~Povh, Phys. Lett. {\bf B245}
(1990) 653
\bibitem{e705} The E705 Collaboration, L.~Antoniazzi et al.
Phys. Rev. Lett. {\bf 70} (1993) 383
\bibitem{kl} B.Z.~Kopeliovich and L.I.~Lapidus, Sov. Phys. JETP
Lett. {\bf 32} (1980) 592
\bibitem{jk} B.K.~Jennings and B.Z.~Kopeliovich, Phys. Rev. Lett.
{\bf 70} (1993) 3384
\bibitem{bm} S.J.~Brodsky and A.~Mueller, Phys. Lett. {\bf B206}
(1988) 685
\bibitem{kz} B.Z.~Kopeliovich and B.G.~Zakharov, Phys. Rev. {\bf
D44} (1991) 3466
\bibitem{bkmnz} O.~Benhar, B.Z.~Kopeliovich, Ch.~Mariotti,
N.N.~Nikolaev and B.G.~Zakharov, Phys. Rev.  Lett.  {\bf 69}
 (1992) 1156
\bibitem{satz} H.~Satz, In "Aachen 1992, Proceedings, QCD: 20
Years Later", v. 2, p. 748

\end{thebibliography}
\end{document}